%% file: main.tex
\documentclass[letterpaper,twocolumn,10pt]{article}
\usepackage{usenix}
\usepackage{balance}
\usepackage{booktabs}
\usepackage{graphicx}
\usepackage{tabularx}
\usepackage{amsmath}
\usepackage{placeins}
\usepackage{tikz}
\usepackage[ruled,vlined,linesnumbered]{algorithm2e}
\usepackage{xspace}
\usetikzlibrary{arrows.meta,positioning}
\newcommand{\cxi}{\textsc{CXI}\xspace}
\newcommand{\contextcxi}{Context-to-Execution Integrity\xspace}

\newcommand{\wxr}{\cxi}
\newcommand{\contextwxr}{\contextcxi}

\begin{document}

\date{}

\title{\Large \bf Context-to-Execution Integrity for LLM Agents}

\author{
{\rm Igor Santos-Grueiro}\\
International University of La Rioja}
\maketitle

\begin{abstract}
\input{sections/00_abstract}
\end{abstract}

\input{sections/01_introduction}
\input{sections/02_problem_threat_model}
\input{sections/03_design}

\input{sections/04_implementation}
\input{sections/05_evaluation_results}
\input{sections/06_discussion_limitations}
\input{sections/07_related_work}
\input{sections/08_conclusion}
\input{sections/08_ethics_open_science}

\bibliographystyle{plain}
\bibliography{references}

\appendix
\balance
\input{sections/09_appendix}

\end{document}

%% file: sections/00_abstract.tex
Language-model agents read attacker-writable context to solve tasks. Tool
execution needs a separate authority check for protected sink fields,
sink-interpreted payloads, and the invocation event. \contextwxr (\wxr) is an
execution-boundary system for this setting. Policies mark protected sink
fields, typed releases carry narrow validated values from writable context to
specific destinations, opaque data slots keep evidence as data, and a
deterministic gate admits a call only after field authority, exact-effect
authorization, and invocation authority all bind to the same action manifest.

We evaluate \wxr on open-weight field-projection runs, AgentDojo live
episodes, a code-agent exact-effect benchmark, manifest-bound ledger faults,
proposal-pressure controls, and hosted/API compatibility traces. AgentDojo
covers 720 live episodes and 1,739 LLM calls; the code-agent benchmark covers
400 repository episodes with
exact-effect authorization and lease-bound execution, yielding 231 safe task
completions and zero observed field, effect, or invocation escapes. The
accounting reports parser outcomes, authorization outcomes, and task-quality
outcomes together with the admission-integrity result. Across the evaluated
sinks, \wxr admits execution only when field, effect, and invocation authority
bind to the same action manifest.

%% file: sections/01_introduction.tex
\section{Introduction}

Language-model agents now turn text into side effects: they send mail, modify
files, update calendars, open pull requests, run shell commands, query
databases, and call administrative APIs~\cite{react,toolformer,webgpt}. They
also read emails, webpages, issue comments, README files, retrieval chunks, CI
logs, tool outputs, package metadata, and memories that an attacker can write
or influence~\cite{greshake2023indirect,agentdojo,toolemu,owasp-llm01}. Many
agent stacks put those sources in the same model context that proposes the next
tool call.

The resulting failure is not just that malicious text may be followed as an
instruction. The systems failure is \emph{authority laundering}: a value that
should be evidence gains authority to select, authorize, trigger, or
parameterize a privileged side effect. A CI log may identify the file that
failed; it should not choose \texttt{prod\_admin.run\_sql}, mark a change as
approved, expand a retry budget, or trigger a second production write. An issue
body may explain a bug; it should not rewrite CI policy or delegate to a more
privileged agent.

Consider a structured tool call with a tool name, operation, path, approval
state, and retry budget. The call may be well formed even when those values come
from writable context. A schema checks names and types; it does not check
authority for the protected fields, interpreted effect, or invocation.

The security question is then whether the
\emph{invocation} has authority to occur and whether each protected field has
authority for the semantic value consumed by the sink. A sink is the
side-effecting endpoint: a tool call, API call, file write, shell command,
database operation, workflow update, or delegation step. A \texttt{FilePath}
typed release may authorize \texttt{file\_path}. The same log or issue text has
no authority for \texttt{approval\_state}. The failing log line may still be
copied as opaque evidence in a comment. The model proposes; the host admits
execution only when the invocation, protected fields, and sink-interpreted
effects carry authority for the same canonical action.

Recent work has moved agent security toward action boundaries. PACT shows that
whole-call trust is too coarse and that argument-level provenance can recover
utility while blocking untrusted content from binding authority-bearing
arguments~\cite{pact2026}. ARGUS audits decisions with influence
provenance~\cite{argus2026}; FORGE develops runtime policies for agentic
systems~\cite{forge2026}; and AIRGuard frames defense as action-time authority
control~\cite{airguard2026}. \wxr builds on this direction but makes a narrower
admission claim for structured side-effect sinks. Plausible field values are
not enough. Selection authority, effect authority, and event authority must all
bind to the same canonical action before execution.

\contextwxr (\wxr) is an execution-boundary system for that
boundary. Its open-weight path generates protected selection fields under
backend evidence; its effect path requires exact authorization for
sink-interpreted payloads; and its execution path consumes an invocation
capability before the sink runs. A deterministic gate ties those obligations to
one canonical action manifest. The model may inspect writable context, but
writable context becomes executable authority only through a destination-scoped
typed release, an exact-effect authorization, or a consumed invocation
capability for the same manifest-bound action.

\wxr is enforced at the last point before the side effect. The gate
canonicalizes the proposed action, checks field evidence, verifies exact-effect
authorizations for payloads
interpreted by the sink, validates semantic bindings under the trusted sink
snapshot, consumes the selected invocation capability in a linearizable ledger,
and executes only with the resulting lease. \wxr's claim is mediated admission
integrity: before a side effect occurs, the host checks that the invocation,
protected fields, and sink-interpreted effects are authorized for the same
canonical action. Task quality, validator completeness, provider internals,
external exactly-once delivery, and complete transformer-level noninterference
are separate obligations.

We evaluate \wxr on open-weight field-projection paths, AgentDojo live
episodes, live code-agent exact-effect tasks, manifest-bound ledger faults,
proposal-pressure controls, and hosted/API compatibility rows. The evaluation
keeps proposals, admissions, leases, external effects, and task success as
separate units. Unauthorized proposals measure attack pressure; field, effect,
and invocation escapes measure whether unsafe authority crossed the mediated
sink boundary; parse failures and fail-closed behavior are not counted as task
success.

This paper makes three contributions:
\begin{itemize}
\item We define \emph{context-to-execution integrity (CXI)}: protected selections,
  sink-interpreted effects, and invocation events must each carry authority for
  the same canonical action before a model proposal becomes executable.
\item We design a manifest-bound admission system that composes
  protected-field evidence, exact-effect authorization, and consumable
  invocation capabilities into one sink decision.
\item We evaluate \wxr across open-weight field-projection paths,
  AgentDojo live episodes,
  a live code-agent exact-effect benchmark, manifest-bound ledger faults,
  proposal-pressure controls, and hosted/API compatibility traces while keeping
  parse failures, utility loss, and validator completeness outside the core
  admission claim.
\end{itemize}

%% file: sections/02_problem_threat_model.tex
\section{Boundary and Threat Model}
\label{sec:problem}

\begin{figure*}[t]
\centering
\begin{tikzpicture}[
  font=\scriptsize,
  source/.style={draw=black!30, rounded corners=2pt, align=left,
    inner sep=3pt, text width=3.15cm, fill=white},
  trusted/.style={source, fill=blue!6, draw=blue!45!black},
  writable/.style={source, fill=red!5, draw=red!45!black},
  release/.style={draw=green!45!black, rounded corners=2pt, align=center,
    inner sep=3pt, text width=2.05cm, fill=green!7},
  opaque/.style={release, draw=black!35, fill=gray!12},
  field/.style={draw=orange!55!black, rounded corners=2pt, align=left,
    inner sep=3pt, text width=3.20cm, fill=orange!7},
  verdict/.style={draw=black!25, rounded corners=2pt, align=center,
    inner sep=2.5pt, text width=1.35cm, fill=white},
  accept/.style={verdict, draw=green!50!black, fill=green!8},
  reject/.style={verdict, draw=red!55!black, fill=red!7},
  dataonly/.style={verdict, draw=black!35, fill=gray!12},
  arr/.style={-{Latex[length=1.7mm]}, line width=.55pt, black!75,
    shorten <=1pt, shorten >=1pt},
  bad/.style={-{Latex[length=1.7mm]}, line width=.65pt, dashed,
    red!70!black, shorten <=1pt, shorten >=1pt},
  data/.style={-{Latex[length=1.7mm]}, line width=.55pt, dotted,
    black!60, shorten <=1pt, shorten >=1pt},
  label/.style={font=\small\bfseries, anchor=west}
]
\node[label] at (-7.05,2.05) {Sources / releases};
\node[label] at (1.25,2.05) {Candidate field};
\node[label] at (5.72,2.05) {Gate};

\node[trusted] (t) at (-5.45,1.25)
  {\textbf{T: registry/task}\\\texttt{repo.write\_file}};
\node[field] (tool) at (2.50,1.25)
  {\textbf{tool/operation}\\\texttt{repo.write\_file / modify}};
\node[accept] (ta) at (6.22,1.25) {ACCEPT};

\node[writable] (wpath) at (-5.45,.35)
  {\textbf{W: CI log}\\\texttt{users.legacy\_token}};
\node[release] (dpath) at (-1.45,.35)
  {\textbf{D(FilePath)}\\typed release};
\node[field] (path) at (2.50,.35)
  {\textbf{file\_path}\\accepted for this field};
\node[accept] (pa) at (6.22,.35) {ACCEPT};

\node[writable] (wquote) at (-5.45,-.55)
  {\textbf{W: log quote}\\diagnostic evidence};
\node[opaque] (opq) at (-1.45,-.55)
  {\textbf{OpaquePayload}\\data only};
\node[field] (comment) at (2.50,-.55)
  {\textbf{comment/evidence}\\quoted for review};
\node[dataonly] (oa) at (6.22,-.55) {DATA\\ONLY};

\node[writable] (wraw) at (-5.45,-1.45)
  {\textbf{W: runbook}\\\texttt{approved}; \texttt{prod\_admin}};
\node[field] (approval) at (2.50,-1.45)
  {\textbf{approval/tool}\\raw W-to-X authority};
\node[reject] (ra) at (6.22,-1.45) {REJECT};

\draw[arr] (t) -- (tool);
\draw[arr] (tool) -- (ta);
\draw[arr] (wpath) -- (dpath);
\draw[arr] (dpath) -- (path);
\draw[arr] (path) -- (pa);
\draw[data] (wquote) -- (opq);
\draw[data] (opq) -- (comment);
\draw[data] (comment) -- (oa);
\draw[bad] (wraw) -- node[midway, above=2pt, fill=white,
  text=red!70!black] {no release} (approval);
\draw[bad] (approval) -- (ra);
\end{tikzpicture}
\caption{Authority laundering at a structured sink. Writable evidence may be
read, quoted, or converted into a typed release for one destination, but raw
writable context cannot authorize protected sink fields such as tool selection
or approval state.}
\label{fig:authority-collapse}
\end{figure*}
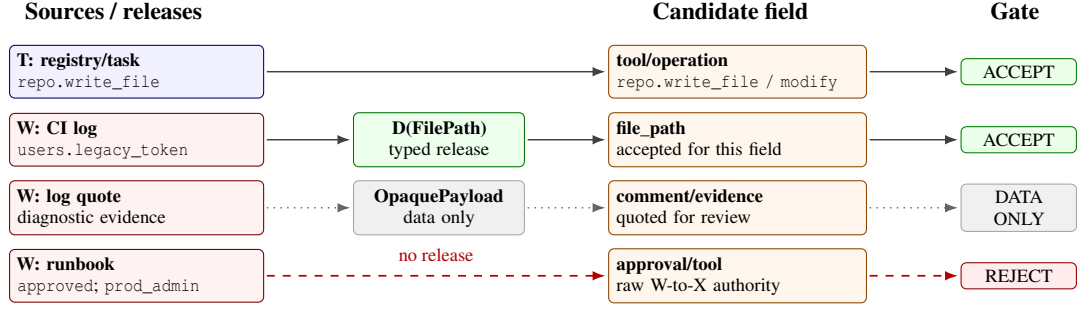

\subsection{Boundary Objects and Fields}

\wxr uses a small vocabulary at side-effect boundaries. In
Figure~\ref{fig:authority-collapse}, the CI log and runbook are readable
evidence, but only trusted state, typed releases, or opaque data slots
can cross the boundary with the authority shown in the verdict column. Trusted
state ($T$) contains the objective, policy, registry, approvals, and other
trusted inputs. Writable context ($W$) is attacker-writable or
attacker-influenced input. Typed releases ($D$) carry validated values from $W$
to a specific destination field. Opaque data ($O$) is copied untrusted text
with no current-boundary authority. Protected sink fields ($X$) select,
authorize, or parameterize an external effect.

\noindent\textbf{Vocabulary.}
CXI distinguishes five kinds of state at the boundary.
\begin{center}
\scriptsize
\setlength{\tabcolsep}{3pt}
\begin{tabularx}{0.98\columnwidth}{l l X}
\toprule
Term & Symbol & Meaning \\
\midrule
Trusted state & $T$ & User objective, policy, registry, approvals, and other
trusted authority sources. \\
Writable context & $W$ & Attacker-writable or attacker-influenced input read by
the agent. \\
Typed release & $D$ & Typed release from $W$ to a specific
destination field. \\
Opaque data & $O$ & Copied $W$ data without authority at the current
boundary. \\
Protected sink field & $X$ & Sink field that selects, authorizes, or parameterizes
a side effect. \\
\bottomrule
\end{tabularx}
\end{center}
Policy tables may group protected sink fields and effect-bearing fields under a
\textsc{HighIntegrity} enum; the main text names each case directly.

Sink schemas partition fields by authority. Every field in a mediated sink must
have an explicit class; unknown fields and omitted classes are rejected.
Protected sink fields include tool, operation, approval, recipient, namespace,
target, path, CI target, delegation, package-install target, and any argument
that can change the external effect. Declared non-effect data is allowed only
after the side effect is authorized and only when the adapter demonstrates that
the sink treats the field as inert. If a field can trigger renderers, slash
commands, webhooks, bots, downstream agents, policy interpretation, or later
side-effecting workflows, it is opaque data that re-enters as $W$ or a
protected or effect-bearing field. Opaque fields may contain uninterpreted
$W$ payloads, so the same string can be admissible in a comment and
inadmissible in an approval field.

Some payloads look like data but still carry effect authority. Patch diffs,
file contents, shell text, SQL, CI YAML, package manifests, policy files,
deployment configuration, generated code, and install hooks can change behavior.
\wxr treats them as effect-bearing unless a policy-specific validator
constrains their destination, scope, and allowed effect.

Scopes keep release decisions local. A \texttt{FilePath} extracted from CI may
authorize \texttt{arguments.file\_path}; it cannot authorize \texttt{tool},
\texttt{operation}, \texttt{approval\_state}, or later package-install targets.
A typed release is therefore scoped to one destination field.

\subsection{Threat Model}

The attacker controls writable-context sources but not trusted policy, the
explicit user objective, model weights, host runtime, or privileged tools. The
attacker may write issues, documentation, CI logs, tool outputs, retrieval
documents, package metadata, persistent memories, or another agent's messages.
The goal is to influence a privileged sink: select a tool, change an operation,
mark approval as satisfied, choose a recipient or namespace, skip confirmation,
persist false authority, or delegate to a more privileged agent.

\contextwxr mediates one boundary: whether adversarially writable data has
authority to influence privileged side-effect fields. It composes with normal
authorization, sandboxing, host integrity, and service-specific access control.
The trusted components are explicit: policy, provenance tracking,
declassifier validators, backend evidence verification, and the host runtime.

\paragraph*{Non-goals.}
\contextwxr relies on a trusted host runtime, trusted tools, correct approval
state, and mediated sinks. It checks field authority before execution. Semantic
task quality remains a separate concern: a patch, diagnosis, or workflow can be
wrong even when its side-effect fields have valid authority. Opaque text can
influence humans, bots, or later agents outside the current mediated sink; when
it re-enters a side-effecting component, it is treated as writable context
again.

\subsection{Provenance and Security Goal}

\wxr tracks provenance of influence, not only provenance of bytes. A field such
as \texttt{tool="repo.write\_file"} may be spelled with bytes from a trusted
registry while the choice to call it came from writable text; that field is
W-derived for \wxr. Each protected field has a conservative influence set
$\Pi[f]$, and the invocation event has $\Pi_{\mathrm{call}}$ for whether the
action is emitted, retried, scheduled, ordered, or batched. Unknown, stale,
repaired, or ambiguous provenance rejects.

The security goal is field-level authority admissibility plus invocation
authority for the side-effect event. A protected sink field may be selected
only by trusted authority valid for that field or by a matching typed release.
Raw W may explain a release; it is not itself field or invocation authority.
Constructive payloads such as patches, file bodies, SQL, shell commands, CI
YAML, package manifests, and deployment descriptors use adapter-specific
exact-effect commitments over the object interpreted by the sink under its
trusted snapshot, principal, defaults, policy epoch, and manifest. The
validator decides which effect is acceptable; \wxr binds admission so the
executed effect must match the effect authorized for that adapter and snapshot.

Deployments may produce different evidence records, but they feed the same
sink-admission check. Operational rules for reads, summaries, memory, tool
outputs, schema repair, cached prefixes, and agent handoff follow the same
conservative influence rule. As in classical information-flow work, the
security question is what information could have influenced an output, not
merely where a string originated
from~\cite{denning1976lattice,goguen1982security,volpano1996sound,sabelfeld2003language}.

For every accepted sink call, $\Pi_{\mathrm{call}}$ must authorize the
invocation under the action history, budget, ordering, idempotency, schedule,
and retry policy. Each protected sink field must satisfy the field rule:

\[
  W \not\rightarrow X
  \quad\text{unless}\quad
  W \rightarrow D_{\mathrm{authorized}} \rightarrow X.
\]

The invocation event has the analogous call-level rule:

\[
  W \not\rightarrow \mathrm{Invoke}
  \quad\text{unless}\quad
  W \rightarrow D_{\mathrm{call}} \rightarrow \mathrm{Invoke}.
\]

The claim separates two requirements. \emph{Manifest-bound admission}
is the gate property: given policy, validators, complete mediation, and
conservative field and call provenance, the gate enforces admission before
execution. \emph{CXI-Provenance} is the runtime requirement: the record must
conservatively capture the influences on each protected field and on the
invocation event. Open-weight and hosted/API deployments use different
records for this purpose; neither claims end-to-end model independence.

This use of typed releases follows prior IFC systems: declassification is an
explicit downgrade boundary with a policy justification
~\cite{sabelfeld2009declassification,myers1999jflow,myers2000dlm,fides,pfi}.

%% file: sections/03_design.tex
\section{Design}
\label{sec:design}

\subsection{Boundary Overview}

\contextwxr checks authority at the action boundary. A model may propose a
structured action, but the host admits it only through ordered boundary steps.

\noindent\textbf{Admission pipeline.}
\cxi admits an action through six ordered boundary steps: (1) \textbf{canonicalize}
the candidate into an action manifest; (2) \textbf{check fields} against
field-local authority and provenance; (3) \textbf{check effects} against
exact-effect commitments; (4) \textbf{authorize invocation} with a
manifest-bound capability; (5) \textbf{consume} that capability to obtain an
execution lease; and (6) \textbf{execute} only through the mediated sink.

A single JSON object may contain a tool name, operation, approval state,
recipient, path, command intent, and payload, but each field has its own
authority requirement. The manifest binds those requirements to one canonical
action so that a release for one field cannot authorize another field, an
effect authorization cannot be replayed against another sink state, and an
invocation capability cannot be spent on a different action.

The design combines total field policy, typed releases, opaque data slots,
field/effect gate checks, manifest-bound invocation capability, and
deployment-specific provenance records. This follows complete mediation and
least privilege: privileged side effects are checked at the boundary, and
authority is never inferred from text
content alone~\cite{saltzer-schroeder,hardy1988confused,watson2010capsicum}.

\begin{table*}[t]
\centering
\scriptsize
\setlength{\tabcolsep}{4pt}
\caption{Authority objects in the main CXI boundary. The model can propose
values, but field selection, sink-interpreted effects, and the invocation event
need separate authority bound to the same canonical action manifest.}
\label{tab:authority-object-summary}
\begin{tabularx}{\textwidth}{@{}p{0.17\textwidth}p{0.23\textwidth}p{0.31\textwidth}X@{}}
\toprule
Object & Authority question & Accepted source & Rejected laundering \\
\midrule
Tool/operation &
Which sink action? &
Trusted registry and task policy. &
Issue or log text chooses the tool. \\
Target path &
Which target value? &
Scoped \texttt{FilePath} release for this field. &
The same path chooses a CI target or command. \\
Patch or file body &
Which interpreted effect? &
Exact-effect authorization over the state delta. &
Raw W patch bytes execute directly. \\
Approval &
Is approval satisfied? &
Trusted approval state or \texttt{ApprovalRef}. &
Copied ``approved'' string from W. \\
Invocation event &
Should the action occur or retry? &
Manifest-bound invocation capability. &
W controls schedule, batch, or retry. \\
Comment/evidence &
Can W be copied as data? &
Opaque slot with preserved provenance. &
Quoted text reused as future authority. \\
\bottomrule
\end{tabularx}
\end{table*}

\subsection{Policy Model}

A CXI policy is the trusted authority map for a mediated sink. It declares
writable origins, trusted roots, mediated sinks, total field classifications,
allowed declassifiers, opaque slots, effect validators, backend evidence
requirements, and authorized sink operations. The policy is fail-closed:
unknown fields, omitted classifications, invalid branch expansions, stale
policy versions, and unsupported evidence modes reject before execution.
Writable text may describe an emergency exception, a new namespace, or a
policy change. Those claims become authority only if trusted state or a
policy-authorized declassifier grants that authority.

The policy is field-scoped. A \texttt{FilePath} release may authorize
\texttt{arguments.file\_path}; it cannot choose \texttt{tool},
\texttt{operation}, \texttt{approval\_state}, a shell command, or a CI target.
Table~\ref{tab:authority-object-summary} is the compact design rule: the
manifest binds authority to field paths, interpreted effects, and invocation
events, not to surrounding natural-language text.

Policies also classify effect-bearing payloads. Patch diffs, file bodies, shell
commands, SQL statements, CI workflows, package manifests, policy files,
deployment descriptors, and install scripts are effect-bearing unless a
validator bounds their effect. Opaque slots remain useful for quoting evidence
or writing user-visible descriptions; they carry data only.
Declared non-effect data is valid only when the adapter demonstrates that the
field is inert; anything that can affect the sink or a downstream actor is
protected, effect-bearing, or opaque with reentry as W.

\subsection{Declassification and Opaque Data}

A declassifier is trusted code that mints a narrow typed release. A model or
extractor may propose a path, approval reference, dependency name, time
interval, or patch intent, but its output remains W until trusted declassifier
code canonicalizes the candidate, validates it against the trusted snapshot,
and mints a typed release (D) scoped to a destination field. The release
carries its output type, source scope, destination scope, declassifier
identity, validator identity, and provenance.
Valid releases are narrow: a normalized time interval, a repository-relative
path, a bounded enum, a hash, a trusted approval identifier, or an opaque
payload handle. Broad natural-language rationales stay data, and syntactically
valid values still reject when outside policy. Representative releases include
\texttt{FilePath}, \texttt{ApprovalRef}, and \texttt{OpaquePayload}.

Opaque data handles preserve evidence without granting authority. A log line,
ticket body, PR description, commit message, or email body can be copied into a
declared evidence field, but the copied body keeps provenance. If a later agent,
bot, tool, or memory system reads it and can create side effects, the payload
enters that boundary as writable context. This follows the declassification
view from IFC: each release states what is released, where it may flow, and
under whose authority~\cite{sabelfeld2009declassification,myers2000dlm,krohn2007flume}.

\subsection{Gate Semantics}

The gate is the only component allowed to turn a schema-valid candidate action
into an executable sink call. It receives the canonical action, a field
provenance map $\Pi$, a call provenance set $\Pi_{\mathrm{call}}$, policy $P$,
effect commitments, and any required backend or field-local evidence records.
It returns either reject or an execution lease. Byte origin alone is
insufficient: a field can contain a trusted-looking string while its selection
was caused by W, and summaries, memory writes, tool outputs, and schema repairs
preserve W influence until a typed release authorizes the destination.

\noindent\textbf{Inputs.}
The core objects are intentionally small. A \emph{PolicyEntry}, keyed by
$(sink, field)$, declares the class, accepted influence atoms, validators,
scopes, policy version, and backend evidence requirement. A canonical
\emph{Value} is interpreted under the sink snapshot. The provenance map $\Pi$
assigns each field a conservative set of $T$, $W$, $D$, and $O$ atoms.
\emph{InvokeCapability} is the matching authority object for the side-effect
event: it names the sink, operation, budget, sequence state, idempotency token,
expiry, trusted snapshot, and policy version, and a linearizable consume must
return a lease before execution. The full call-atom rules, control-flow
propagation rules, capability resolver, and verifier are in the appendix.

\noindent\textbf{Authority predicates.}
\texttt{mayInfluence(P, sink, field, atom)} answers whether an atom may
participate in choosing this field. \texttt{authorizesValue(P, sink, field,
value, atom)} is stricter: it asks whether that atom authorizes the actual
semantic value or effect by exact-value digest, exact-effect commitment, or a
bounded predicate over the canonical value and trusted snapshot. A protected
sink field is accepted only if every atom in $\Pi_f$ may influence the field
and at least one atom authorizes the selected semantic value or effect. W,
opaque, unknown, wrong-domain trusted, cross-field D, unknown optional
branches, and missing referenced fields reject.

\noindent\textbf{Operational field check.}
Operationally, the field predicate checks six conditions. First, the field must
be in the active sink schema and must have a total policy classification.
Second, the field must
carry known conservative provenance. Third, opaque slots may preserve trusted
or opaque data but grant no later authority. Fourth, declared non-effect data
is admitted only for adapter-demonstrated inert fields. Fifth, every influence
on a protected sink field must be trusted or a destination-scoped typed release
(D) for that exact field, and at least one atom must authorize the canonical
value or adapter-specific effect. Sixth, validators and relational action
constraints must accept under the trusted snapshot.

\noindent\textbf{Fail-closed behavior.}
Raw W provenance in a protected sink field, D outside scope, wrong type, opaque
text in an authority field, unauthorized tool or operation, schema repair over
a protected field, backend evidence record mismatch, and unsupported protected
backend paths all make the sink call non-executable. In other words, any
failure to satisfy field, effect, or invocation authority causes the gate to
reject before execution.

\subsection{Exact-Effect Authorization and Action Manifests}
\label{sec:exact-effect-manifest}

Field isolation is sufficient only when the protected object is the value the
sink consumes. Constructive payloads such as patches, file bodies, SQL,
commands, workflows, package manifests, or deployment descriptors may
legitimately depend on W because the task comes from an issue, log, test
failure, or user document. That dependency makes the payload useful, not
authorized. Simple canonical fields use \textsc{ExactValue}$(hash)$ over the
sink-parsed semantic value. Effect-bearing payloads use
\textsc{ExactEffect}$(commitment)$: the adapter computes the effect the sink
will apply under the trusted snapshot, policy epoch, effective principal,
explicit defaults, adapter revision, validator set, and action nonce.

\begin{table}[t]
\centering
\scriptsize
\setlength{\tabcolsep}{3pt}
\caption{Exact-effect authorization binds the sink-interpreted effect, not the
payload spelling. Validators decide which exact effect is acceptable; CXI binds
the accepted effect to the action that may be admitted.}
\label{tab:exact-effect-commitments}
\begin{tabularx}{\columnwidth}{lX}
\toprule
Payload & Authorized commitment \\
\midrule
Repository patch & Repository state delta over base commit, canonical paths,
old/new blobs, modes, symlinks, renames, and allowed path set. \\
File body & Canonical path, bytes, mode, target snapshot, and adapter revision. \\
SQL & Parsed statement or AST, parameters, schema snapshot, role, and allowed
write set. \\
Shell command & Trusted-rendered argv/env/cwd plus sandbox, network, filesystem,
and process policy. \\
CI or package manifest & Canonical state delta plus workflow, dependency,
install-hook, and policy-file predicates. \\
\bottomrule
\end{tabularx}
\end{table}

The action manifest is the final binding object. It records the action
nonce, canonical action digest, adapter ID/revision, effective principal,
trusted snapshot, policy epoch, assembly revision, protected-field evidence
digests, effect-authorization digests, and call-authority digest. The
capability ledger consumes that manifest digest.

Any mutation after validation changes a digest and causes execution to fail.
Replaying the same payload bytes against a different repository, workspace,
base commit, validator set, adapter revision, principal, or policy epoch changes
the manifest and rejects the action. The manifest binds the action, the
field authorizations, the effect authorization, and the invocation capability
into one commitment.

CXI runs only the exact effect the
validator authorized and only the exact manifest the capability admitted. CXI
does not judge whether the effect is desirable; it enforces that the admitted
effect matches the exact effect trusted validator code authorized for this
snapshot, principal, adapter, and policy epoch. Validator completeness remains
a separate adapter obligation.

\subsection{Admitted-Action Contract}

Fix a policy $P$, fail-closed validators, conservative field provenance $\Pi$,
conservative call provenance $\Pi_{\mathrm{call}}$, complete mediation of
privileged sinks, and any backend or field-local evidence required by policy.
If the CXI gate accepts a sink call, the accepted action satisfies four
obligations.

\noindent\textbf{Invocation authority.}
The invocation event is authorized by a valid invocation capability for the
sink, operation, task/run, sequence state, retry budget, idempotency key,
schedule, history, trusted snapshot, policy epoch, and adapter revision. All
call-level influences in $\Pi_{\mathrm{call}}$ may influence that invocation
under policy. The selected capability is the one consumed by the linearizable
capability store, and the store returns an execution lease before the sink
executes.

\noindent\textbf{Field authority.}
Every executed protected sink field has only trusted authority valid for that
field or a policy-valid typed release (D) whose type, source scope, destination
scope, validator, value binding, action context, and policy version match that
field's policy. At least one trusted or D atom authorizes the executed semantic
value by exact digest, exact-effect commitment, or bounded predicate.

\noindent\textbf{Effect binding.}
For any sink-interpreted payload, the executed effect equals the effect
commitment authorized by the adapter for the trusted snapshot, principal,
policy epoch, validator set, adapter revision, and action nonce.

\noindent\textbf{Opaque reentry.}
Opaque data may only go into declared non-authority slots. If it later
enters another agent, tool, bot, memory, or workflow that can create side
effects, that boundary treats it as W again.
Raw W, W-derived summaries or memory, opaque data in authority fields,
missing provenance, stale metadata, repaired protected fields,
evidence record mismatches, and out-of-scope typed releases fail closed. The
gate cannot infer these facts at runtime: sinks must be mediated, policies must
classify fields correctly, provenance must conservatively track influence,
validators must fail closed, and evidence records must match the required path.

\noindent\textbf{Composition with provenance records.}
The admitted-action contract works with different provenance records.
Open-weight deployments can satisfy the field-provenance requirement with
field-closed protected-field decoding; direct-span, suffix-taint, and
runtime-map exclusion records are narrower evidence checks for direct W-source,
generated-prefix, and declared-atom exclusion. Hosted/API deployments satisfy
the requirement with field-local writer visibility under the host. Both
still require call-level provenance for the invocation event.

\subsection{Authority-Projected Protected-Field Decoding}

The open-weight path gives the strongest evidence because the runtime can
constrain protected-field generation before the sink gate sees a completed
tool call. For each protected field $f$, policy builds a field-specific
projection containing only trusted state, typed releases valid for $f$, authorized
parent field records, the field grammar, candidate/value authorizers, backend
profile, action nonce, and sampling configuration. The decoder emits one field
value and one backend evidence record; the host assembles the canonical action
and the gate verifies the record before consuming invocation authority.

\begin{table}[t]
\centering
\scriptsize
\setlength{\tabcolsep}{3pt}
\caption{Field-closed reference contract. This is the reference target for
open-weight protected-field decoding; optimized serving paths must match this
target on deterministic canaries or fail closed.}
\label{tab:field-closed-contract}
\begin{tabularx}{\columnwidth}{lX}
\toprule
Reference requirement & Meaning \\
\midrule
One protected field & Each decode produces one authority-bearing field. \\
Physical projection & The prompt contains only T, field-valid typed releases, and
authorized parent field records for $f$. \\
Clean reference state & Reference mode uses fresh prefill and fresh KV, with no
W-seen prefix reuse, batching, CUDA graph replay, or speculative decode. \\
Field grammar & Grammar and candidate/value predicates are local to $f$. \\
Host assembly & The host assembles fields deterministically after per-field
decodes. \\
Field evidence & The backend evidence record binds the projected input, grammar,
raw output, canonical value, parent records, backend profile, nonce, and action
manifest. \\
\bottomrule
\end{tabularx}
\end{table}

The contract separates the reference target from evidence checks. Direct-span,
suffix-taint, and runtime-map records check declared W source, generated-prefix,
and runtime-map influence; they are not field-closed lineage unless they
satisfy Table~\ref{tab:field-closed-contract}. Hosted/API deployments cannot
observe provider internals, so they use field-local construction evidence
instead. Both paths feed the same manifest-bound sink gate.

\begin{table}[t]
\centering
\scriptsize
\setlength{\tabcolsep}{3pt}
\caption{Two provenance paths for the same sink-authority check. The
open-weight path provides backend evidence for protected-field decoding; the
hosted/API path provides field-local construction evidence when provider
internals are unavailable.}
\label{tab:enforcement-regimes}
\begin{tabularx}{\columnwidth}{lXX}
\toprule
Path & Evidence produced & Boundary checked \\
\midrule
Open-weight & Backend evidence record with mask state, blocked ranges, action digest, field path, and evidence record serialization mode &
Serving stack plus final sink gate \\
Hosted/API & Extractor record, field-local writer visibility, typed releases,
opaque data handles, final action provenance &
Host plus final sink gate \\
\bottomrule
\end{tabularx}
\end{table}

Open-weight and hosted/API paths differ in the evidence they can produce, not
in the sink-authority check. Section~\ref{sec:evaluation} reports the
evaluated surfaces.

%% file: sections/04_implementation.tex
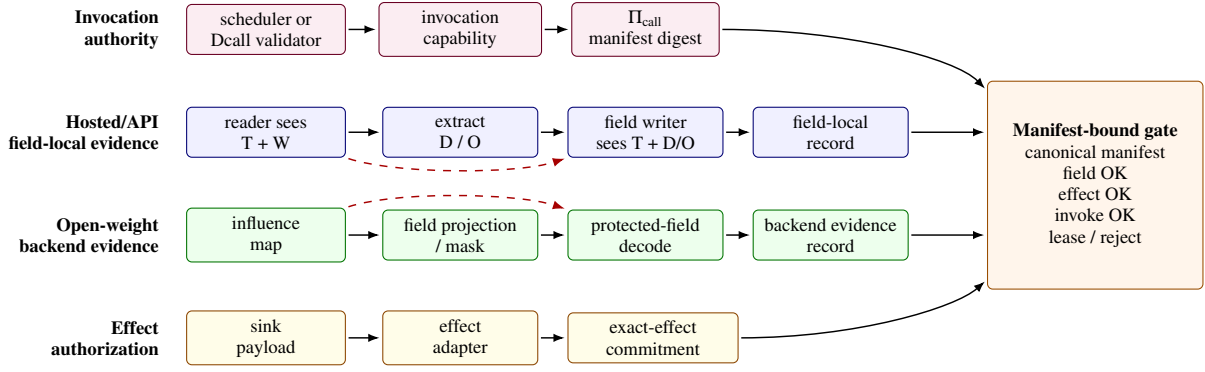
\begin{figure*}[t!]
\centering
\scriptsize
\begin{tikzpicture}[
  font=\scriptsize,
  box/.style={draw, rounded corners=2pt, align=center, inner sep=3pt,
    minimum height=.66cm, fill=gray!6},
  inv/.style={box, fill=purple!7, draw=purple!55!black},
  api/.style={box, fill=blue!6, draw=blue!45!black},
  ow/.style={box, fill=green!7, draw=green!45!black},
  eff/.style={box, fill=yellow!10, draw=orange!55!black},
  gate/.style={box, fill=orange!8, draw=orange!60!black,
    minimum height=2.75cm},
  lane/.style={font=\scriptsize\bfseries, anchor=east, align=right},
  arr/.style={-{Latex[length=1.5mm]}, line width=.55pt, shorten <=1pt, shorten >=1pt},
  barr/.style={-{Latex[length=1.6mm]}, line width=.55pt, dashed, red!65!black,
    shorten <=1pt, shorten >=1pt}
]
\node[lane] at (-6.25,2.28) {Invocation\\authority};
\node[lane] at (-6.25,.92) {Hosted/API\\field-local evidence};
\node[lane] at (-6.25,-.44) {Open-weight\\backend evidence};
\node[lane] at (-6.25,-1.80) {Effect\\authorization};

\node[inv, minimum width=2.10cm] (scheduler) at (-4.92,2.28) {scheduler or\\Dcall validator};
\node[inv, minimum width=2.15cm] (cap) at (-2.35,2.28) {invocation\\capability};
\node[inv, minimum width=1.95cm] (callrec) at (.10,2.28) {$\Pi_{\mathrm{call}}$\\manifest digest};

\node[api, minimum width=2.10cm] (wapi) at (-4.92,.92) {reader sees\\T + W};
\node[api, minimum width=2.05cm] (extractor) at (-2.35,.92) {extract\\D / O};
\node[api, minimum width=2.05cm] (writer) at (.10,.92) {field writer\\sees T + D/O};
\node[api, minimum width=2.05cm] (apirec) at (2.55,.92) {field-local\\record};

\node[ow, minimum width=2.10cm] (tow) at (-4.92,-.44) {influence\\map};
\node[ow, minimum width=2.05cm] (allowed) at (-2.35,-.44) {field projection\\ / mask};
\node[ow, minimum width=2.05cm] (decode) at (.10,-.44) {protected-field\\decode};
\node[ow, minimum width=2.05cm] (evidence) at (2.55,-.44) {backend evidence\\record};

\node[eff, minimum width=2.10cm] (payload) at (-4.92,-1.80) {sink\\payload};
\node[eff, minimum width=2.05cm] (adapter) at (-2.35,-1.80) {effect\\adapter};
\node[eff, minimum width=2.25cm] (commitment) at (.20,-1.80) {exact-effect\\commitment};

\node[gate, minimum width=2.85cm] (gate) at (6.05,.22)
  {\textbf{Manifest-bound gate}\\canonical manifest\\field OK\\effect OK\\invoke OK\\lease / reject};

\draw[arr] (scheduler) -- (cap);
\draw[arr] (cap) -- (callrec);
\draw[arr] (callrec.east) .. controls (2.75,2.28) and (4.20,1.90) .. ([yshift=1.25cm]gate.west);

\draw[arr] (wapi) -- (extractor);
\draw[arr] (extractor) -- (writer);
\draw[arr] (writer) -- (apirec);
\draw[arr] (apirec.east) -- ([yshift=.70cm]gate.west);

\draw[arr] (tow) -- (allowed);
\draw[arr] (allowed) -- (decode);
\draw[arr] (decode) -- (evidence);
\draw[arr] (evidence.east) -- ([yshift=-.66cm]gate.west);

\draw[arr] (payload) -- (adapter);
\draw[arr] (adapter) -- (commitment);
\draw[arr] (commitment.east) .. controls (2.90,-1.80) and (4.20,-1.40) .. ([yshift=-1.25cm]gate.west);

\draw[barr] (wapi.south east) to[out=-18,in=198,looseness=.62] (writer.south west);
\draw[barr] (tow.north east) to[out=18,in=162,looseness=.62] (decode.north west);
\end{tikzpicture}
\caption{CXI admission architecture. Invocation, hosted/API, open-weight, and
effect-adapter lanes produce separate authority or evidence records for one
canonical manifest. The manifest-bound gate verifies field authority,
exact-effect authorization, and invocation capability before issuing a lease.
Dashed paths show writable influence excluded from protected-field construction.}
\label{fig:cxi-architecture}
\end{figure*}

\section{Implementation}

CXI is implemented as one shared sink gate with deployment-specific provenance
records. The open-weight path gives the strongest evidence because the runtime
can constrain protected-field generation directly. The hosted/API path is for
models whose attention masks and cache state are unavailable: it records what a
reader and controlled field constructors were allowed to see. Both paths end at
the same execution point: execute only when the invocation event, protected
fields, and sink-interpreted effects are authorized for the same canonical
manifest.

\subsection{Common Execution Boundary}

A policy is a deterministic data object naming writable origins, trusted roots,
mediated sinks, field classes, opaque slots, declassifiers, effect validators,
evidence requirements, and allowed sink operations. The gate classifies
authority, not text intent: it canonicalizes the candidate action, attaches
field influence records, rejects unmediated sinks or invalid policies,
authorizes the invocation event, validates protected fields, verifies backend
evidence records when required, consumes the
capability, and executes only with the returned lease.

Complete mediation is required: a CXI result is meaningful only for
side-effect paths that enter the shared gate. Every effect interface must
be routed through this manifest-bound control point, and alternate paths
must be blocked. Table~\ref{tab:main-mediation} lists representative entry
points and negative bypass checks.

\begin{table*}[t]
\centering
\scriptsize
\setlength{\tabcolsep}{3pt}
\caption{Representative complete-mediation checks. CXI's admission result
is meaningful only for side-effect paths routed through the shared
manifest-bound gate.}
\label{tab:main-mediation}
\begin{tabularx}{\textwidth}{@{}p{0.12\textwidth}p{0.24\textwidth}p{0.30\textwidth}X@{}}
\toprule
Sink family & Mediated entry point & Authority checked & Negative bypass check \\
\midrule
Filesystem & Write, delete, move, and generated-file operations. &
Path, operation, payload effect, trusted snapshot, and adapter revision. &
Direct file APIs and helper paths are denied or wrapped. \\
Shell/process & Trusted-rendered argv/env/cwd plus sandbox policy. &
Command intent, environment, cwd, network, filesystem, and process scope. &
Raw shell strings from W do not execute. \\
Network/API & Endpoint, recipient, namespace, method, and idempotency. &
Operation authority, effective principal, endpoint scope, and call lease. &
Alternate clients require the same policy and lease. \\
Repository & Patch, branch, PR, workflow, and CI changes. &
Allowed paths, base commit, old/new blobs, modes, and state delta. &
Patch bytes cannot execute raw or replay on another base. \\
Memory/state & Store, retrieve, summarize, and repair events. &
Influence provenance, freshness, parent records, and destination scope. &
Unlabeled or stale memory reads fail closed. \\
Delegation/bots & Agent target, task, permissions, and event capability. &
Delegation target, task scope, permissions, and invocation authority. &
Opaque data reentry is treated as W. \\
Package/CI & Dependency, registry, workflow, and install-hook changes. &
Registry scope, workflow privilege, policy files, and exact-effect validator. &
Install-hook and workflow side effects require adapter authority. \\
\bottomrule
\end{tabularx}
\end{table*}

Invocation authority is checked separately from argument authority. The
predicate validates sink, operation, precondition, sequence state, retry count,
batch cardinality, idempotency key, budget, expiry, task/run ID, and history
against an invocation capability. The reference ledger consumes against the
\texttt{ActionManifest}, not a free-form proposal digest; the capability must
match the manifest's call-authority digest, policy epoch, trusted snapshot,
effective principal, adapter ID, and adapter revision. The implementation
records local consume, deduplication, rejection, and double-spend checks; the
evaluation reports those rows separately. Exactly-once external application
remains an adapter contract requiring durable delivery and sink-side
deduplication or a shared transactional boundary.

\subsection{Declassifiers and Effect Adapters}

\noindent\textbf{Declassifiers.}
Declassifiers are trusted release points. Each specification declares an
input scope, output type, destination field set, canonicalizer, validator, and
negative tests. A model may propose a candidate; trusted policy code mints the
typed release (D) only after validation. Validators are destination-specific: a
\texttt{FilePath} release can authorize \texttt{arguments.file\_path}, not
\texttt{tool}, \texttt{operation}, \texttt{approval\_state}, or a CI target.
Values are normalized under the trusted snapshot: paths resolve against the
repository snapshot, package coordinates against the trusted registry view,
recipients and URLs are canonicalized, and closed command or SQL intents are
rendered by trusted code. Validator families, an auditable \texttt{FilePath}
fragment, and declassifier trust assumptions are summarized in the appendix.

\noindent\textbf{Effect adapters.}
For effect-bearing repository patches, the implementation instantiates
Section~\ref{sec:exact-effect-manifest} with a repository state-delta adapter.
The delta binds repository and workspace identity, base commit, canonical
paths, file modes, old/new blob digests, symlink targets, adapter revision,
validator set, policy epoch, and action nonce. The adapter rejects traversal,
forbidden paths, symlink escapes, CI disablement, install hooks, secret writes,
network side-effect markers, and replay across another nonce, repository,
workspace, base commit, validator set, or adapter revision. The exact-effect
adapter rows are deterministic checks; live code-agent task success is measured
separately.

\noindent\textbf{Opaque data handles.}
Opaque data handles preserve evidence without granting field authority and
re-enter any later side-effecting boundary as writable context.

\subsection{Open-Weight Protected-Field Path}

The open-weight path labels a protected field before it is decoded.
The runtime maps prompt spans to token ranges, labels influence atoms, assigns
the next structured-output field to a policy class, and creates a contract
naming the field path, blocked authority ranges, and any trusted or
typed-release regions allowed for that field. The reference implementation
\texttt{decode\_protected\_field} constructs the field projection, rejects raw
W, opaque data, cross-field releases, wrong-domain authority,
undeclared parents, and unsupported serving paths, and emits a backend
evidence record only for a completed field. This reference implementation is
the reference target for optimized serving paths; open-weight utility and
hosted/API compatibility are evaluated separately.

Optimized paths reuse the same projection and evidence record schema while
adding lineage-compatible cache keys, batch-slot metadata checks,
repeated-field nonce rejection, and fail-closed preflight for CUDA graph replay
and speculative decoding. The evaluated open-weight serving implementation
integrates with vLLM's paged KV-cache
runtime~\cite{vllm} and attention
optimizations~\cite{dao2022flashattention,orca,sglang,speculative-decoding};
the Triton path sets attention scores from protected X query tokens to blocked
W-derived key tokens to $-\infty$ before softmax. An active mask on an
unsupported backend path fails before execution. Table~\ref{tab:field-closed-contract}
gives the reference target and evidence checks: direct-span, suffix-taint, and
runtime-map rows check direct W-source, generated-prefix, and declared active
K/V influence-map exclusion rather than full field-closed model independence.

\subsection{Evidence Verification and Hosted/API Compatibility}

\noindent\textbf{Backend evidence verification.}
A protected-field backend evidence record binds evidence to one field in one
action. The verifier recomputes the canonical action digest, checks freshness
and field binding, confirms a supported backend path, enforces the policy
minimum evidence mode, checks input, influence-map, cache-lineage, and
trusted-snapshot bindings, and then runs the normal gate. Anti-downgrade
rejects direct-span evidence when policy requires suffix-taint, runtime-map, or
field-closed evidence. Any repair, retry, or edit to a protected field changes
the digest and requires a fresh protected-field
decode~\cite{abadi2005cfi,wahbe1993sfi,yee2009nativeclient}. These records
prove the backend path and protected-field binding used for admission; remote
attestation against a compromised host would require a separate mechanism.

\noindent\textbf{Hosted/API compatibility.}
Hosted APIs expose neither attention masks nor KV-cache labels, so this path
cannot rely on provider internals for exclusion. Instead, it records a
field-local construction boundary: an extractor may read the trusted objective
and writable context, but it emits only typed facts, provenance records, and
opaque data handles. Protected fields are then assembled only from trusted
policy and valid typed releases or by a field-specific writer whose visible
inputs are trusted state plus typed releases for that destination. The API
record documents which inputs the host controlled, while provider-internal
decoder dependencies remain inside the hosted provider. Serving optimizations
are allowed only when they preserve CXI metadata.

%% file: sections/05_evaluation_results.tex
\section{Evaluation and Results}
\label{sec:evaluation}

\subsection{Methodology}

\noindent\textbf{Endpoint.}
The evaluation tests an admission boundary, not a single benchmark score. The
primary security endpoint is unauthorized execution: a protected-field value,
invocation event, or sink-interpreted effect reaching the sink without the
corresponding authority. Proposal pressure, canonical-action yield,
benign-call recall, backend evidence validity, and final-state quality are
reported separately; \emph{escape} is reserved for unauthorized execution.

\noindent\textbf{Surfaces and setup.}
The evaluation covers six surfaces: open-weight field projection, AgentDojo
task episodes, code-agent exact-effect tasks, manifest-bound ledger faults,
proposal pressure before the gate, and hosted/API compatibility. Deterministic
rows are fixed checks and fault-injection tests; sampled rows report recorded
behavior. Open-weight sampled runs used NVIDIA H100 80GB HBM3 GPU hosts with
driver 570.172.08, CUDA 12.8, PyTorch 2.11.0+cu128, and vLLM
0.21.1rc1.dev268 precompiled builds; GPT-OSS used native MXFP4. This is the
machine and serving setup, not a result category.

\noindent\textbf{Model matrix.}
Open-weight IDs are \path{Qwen/Qwen3.6-27B},
\path{google/gemma-3-27b-it}, \path{openai/gpt-oss-120b}, and
\path{Qwen/Qwen3-Coder-30B-A3B-Instruct}. Archived hosted/API IDs are
\path{gpt-5.5}, \path{claude-opus-4-7}, \path{claude-sonnet-4-6}, and
\path{gemini-3.5-flash}; hosted aliases are archived record identifiers.
Open-weight vLLM rows use temperature 0.0 and 64 generated tokens; the
appendix records model-specific context, precision, quantization, and hosted/API
parameters.

\noindent\textbf{Accounting.}
One scorer keeps task success, field/effect/invocation integrity,
safe-task-completion, evidence-valid actions, admissions, attempts, effects,
fail-closed outcomes, and false suppressions as separate columns. Each case
fixes trusted task state, writable sources, candidate sink, field classes,
declassifiers, opaque slots, and expected outcome before model output is
scored; records include the proposed action, influence sets, gate decision,
model settings, and backend evidence or API-boundary record. Deployment audits
cover 100 utility workflows and 270 typed or opaque steps.

\subsection{Evaluated Surfaces and Denominators}

Table~\ref{tab:main-results} is the main accounting table. It reports the
evaluated surfaces, their units, the evidence produced before execution, and
the boundary outcome observed at the mediated sink. The rows use different unit
types and should not be averaged: AgentDojo and code-agent measure live task
surfaces, the ledger row measures deterministic admission faults, and the
hosted/API row measures archived field-local compatibility. Safe task completion is
meaningful only for live task episodes; the integrity columns are the admission
boundary.

\begin{table*}[t]
\centering
\scriptsize
\setlength{\tabcolsep}{3pt}
\caption{Main evaluated surfaces and denominators. Rows use different units
and are not averaged. Task utility is meaningful only for live episodes; integrity
outcomes refer to mediated admission.}
\label{tab:main-results}
\begin{tabularx}{\textwidth}{p{2.45cm}p{1.55cm}p{1.15cm}Xp{2.45cm}p{2.65cm}}
\toprule
Surface & Unit / N & Task utility & Admission evidence & Boundary outcome & Accounting note \\
\midrule
AgentDojo live task bank &
720 episodes & 199 STC &
1,739 LLM calls; exact-effect subset: 306 field records, 204 leases/effects &
no observed field/effect/invocation escapes in protected admissions &
STC across six live conditions \\
AgentDojo exact-effect admissions &
204 leases & -- &
306 field records; 204 exact-effect authorizations &
0/0/0 field/effect/invocation &
not a standalone STC denominator \\
Code-agent exact-effect benchmark &
400 repo episodes & 231 STC &
exact-effect authorization + manifest verification + capability consume + lease &
0/0/0 field/effect/invocation &
73 parse failures; 61 no effect authorization; 35 task-quality failures \\
Manifest-bound ledger faults &
10 cases / 29 consumes & -- &
19 rejects; 2 deduplicated retries &
0 duplicate admissions/applications &
adapter-contract claim \\
Hosted/API field-local compatibility &
800 archived rows & -- &
85 unsafe selections blocked; 0 parse failures &
0 gate escapes &
no provider-internal decoder evidence \\
\bottomrule
\end{tabularx}
\end{table*}

The table is not a single-score scoreboard. AgentDojo STC uses all six live
conditions, while the AgentDojo field records, leases, and exact-effect
authorizations come only from exact-effect admission conditions. The
code-agent benchmark exercises the exact-effect and manifest design in
Section~\ref{sec:exact-effect-manifest}, reporting safe task completion
separately from parse failures, parse-ok rows without effect authorization,
and authorized executions that still miss the task oracle. These are utility,
parser, validator, or task-quality outcomes, not field/effect/invocation escapes.
For the code-agent benchmark, non-STC decomposes as 73 parse failures, 61
parse-ok rows with no effect authorization, and 35 authorized/executed rows
without task success.

\subsection{Evidence Contributions to Admission}

\noindent\textbf{Evidence ladder.}
The open-weight rows form an evidence ladder rather than interchangeable
claims, following the design hierarchy in Table~\ref{tab:field-closed-contract}.
Field-closed reference decoding is the reference behavior: one protected field,
a physical field-specific projection, fresh prefill/KV in the reference path,
field grammar, and host assembly. The open-weight serving rows then test serving-path and
utility evidence checks against that target. Unmasked shared decoding is an unsafe
control; direct-span and suffix-taint are ablation contracts; runtime-map
exclusion is a declared-atom check; Hosted/API field-local is a compatibility check
rather than provider-internal mask, attention, or KV-lineage evidence.
The model matrix identifies where the evidence was observed;
Table~\ref{tab:ow-mode-roles} identifies what each evidence surface contributes
before the common gate: field, effect, and invocation evidence for the same
admission check.

\begin{table*}[t]
\centering
\scriptsize
\setlength{\tabcolsep}{3pt}
\caption{Evidence contributions to the admission boundary. Each row names the
evidence contributed before the common manifest-bound gate. Different
deployments expose different evidence, but all feed the same
field/effect/invocation admission check.}
\label{tab:ow-mode-roles}
\begin{tabularx}{\textwidth}{p{3.05cm}p{3.10cm}X}
\toprule
Evidence source & Boundary role & What it contributes \\
\midrule
Field-closed reference & Protected-field authority &
Reference target for field-local authority and host assembly. \\
Open-weight serving check & Protected-field authority &
Serving-path check against the field-closed target. \\
Direct-span / suffix-taint / runtime-map & Open-weight evidence checks &
Increasingly stronger exclusion checks over declared W influence before a
protected field is admitted. \\
Hosted/API field-local & Compatibility evidence &
Host-visible construction evidence when provider internals are
unavailable. \\
Exact-effect adapter & Effect authority &
Binds a sink-interpreted payload to the authorized semantic effect. \\
Manifest-bound ledger & Invocation authority &
Consumes the matching capability, suppresses duplicate admissions under the
adapter contract, and issues the execution lease. \\
\bottomrule
\end{tabularx}
\end{table*}

These rows define the evaluated claim boundary; provider internals, validator
completeness, and external exactly-once delivery require separate checks.

CXI-Core focuses on code agents because they combine attacker-writable inputs
with high-impact repository, shell, package, ticket, and CI actions. The
evaluation spans four open-weight models and four hosted/API models; the
appendix records the model and runtime settings and summarizes the evidence
boundaries. Hosted rows are archived compatibility traces.

\noindent\textbf{Evidence details.}
\wxr uses one sink predicate in open-weight and hosted/API settings, but the
evidence differs. Open-weight rows attach backend evidence to protected-field
generation. Hosted/API rows attach host-observed construction evidence:
field writers receive trusted state plus field-valid typed releases, opaque
data stay out of writer context, and the same final gate checks the assembled
action. Availability is reported separately from evidence validity in both
paths.

\noindent\textbf{Open-weight backend evidence.}
\wxr distinguishes the field-closed target from narrower evidence checks.
Field-closed decoding generates each protected field from a field-valid
projection with clean cache lineage and binds the value into the final action
manifest. The open-weight serving rows are checks toward that target: direct-span
exclusion covers declared raw W source keys but not generated prefixes or other
atoms; suffix-taint exclusion covers both raw W keys and generated W-derived
prefixes; and runtime-map exclusion is a backend ablation over declared runtime
influence atoms for cross-field D, opaque, W, and wrong-trusted relay risks.
Full causal taint over internal transformer state requires stronger serving
support. The gate requires a matching backend evidence record for each protected
field.

In the direct-span validation run, each of the four fixed open-weight
model/backend pairs decodes 768 protected fields with no runtime-visible W
source positions exposed to the privileged field; matched unmasked records are
rejected and invalid evidence admissions remain 0.000. Direct-span evidence
covers declared raw W source keys only. Runtime-map rows add declared-atom
checks and fail closed before the sink when canonical action construction or
evidence validation is incomplete. These rows cover declared source and atom
checks; complete KV-lineage independence would require stronger backend
evidence. Detailed backend roles are summarized in the appendix.

\noindent\textbf{Hosted/API compatibility and offline mapping.}
Hosted APIs expose no attention masks, KV-cache labels, or backend evidence
records. The Hosted/API field-local compatibility path therefore checks host
construction: a reader may see raw W, but protected-field writers see
only trusted state and typed releases valid for that field. Across 800 archived
proposals, the host-observed API compatibility path blocks 85 unsafe
selections and records 0 gate escapes and 0 parse failures. This evidence
supports hosted/API compatibility while provider-internal decoder state,
attention masks, and KV lineage remain unavailable.

Offline policy mappings are reported separately from live task success. The
final mapping audit covers CXI-Core 192, AgentDojo 609, and ToolEmu 144
denominators and records no observed unauthorized admissions in the archived
compatibility rows. The broader evaluated task-bank run records 720 episodes,
1,739 live LLM calls, 306 backend field evidence records, 204 leases, 204 exact-effect
authorizations, and 199 safe task completions across all six defense
conditions. Non-STC leased episodes are task-quality, availability, or
model-output outcomes rather than field, invocation, or effect escapes.
Validator completeness beyond the evaluated AgentDojo suite is not claimed.

\subsection{Proposal Pressure Before Admission}

\begin{figure}[!h]
\centering
\begin{tikzpicture}[
  font=\scriptsize,
  rowlabel/.style={anchor=east, align=right, text width=1.75cm},
  track/.style={draw=black!18, fill=gray!7, rounded corners=1pt},
  pressure/.style={draw=red!60!black, fill=red!22, rounded corners=1pt},
  zerobox/.style={draw=green!45!black, fill=green!8, rounded corners=1pt,
    align=center, inner sep=1.5pt, text width=.42cm},
  valuelabel/.style={anchor=west, align=left}
]
\def\barleft{0}
\def\barw{2.85}
\def\barh{.20}
\def\zerox{5.38}
\node[font=\scriptsize\bfseries, align=center] at (1.43,.48)
  {unauthorized proposals\\before gate};
\node[font=\scriptsize\bfseries, align=center] at (\zerox,.48)
  {executed\\escapes};

\node[rowlabel] at (-.25,0) {Utility\\adversarial};
\draw[track] (\barleft,-\barh/2) rectangle (\barw,\barh/2);
\draw[pressure] (\barleft,-\barh/2) rectangle (1.86,\barh/2);
\node[valuelabel] at (3.05,0) {196/300 field};
\node[zerobox] at (\zerox,0) {0};

\node[rowlabel] at (-.25,-.62) {Stress\\rows};
\draw[track] (\barleft,-.62-\barh/2) rectangle (\barw,-.62+\barh/2);
\draw[pressure] (\barleft,-.62-\barh/2) rectangle (2.22,-.62+\barh/2);
\node[valuelabel] at (3.05,-.62) {407/522 field};
\node[zerobox] at (\zerox,-.62) {0};

\node[rowlabel] at (-.25,-1.24) {Hosted/API\\adversarial};
\draw[track] (\barleft,-1.24-\barh/2) rectangle (\barw,-1.24+\barh/2);
\draw[pressure] (\barleft,-1.24-\barh/2) rectangle (.61,-1.24+\barh/2);
\node[valuelabel] at (3.05,-1.24) {85/400 field};
\node[zerobox] at (\zerox,-1.24) {0};

\node[rowlabel] at (-.25,-1.86) {Free-form\\actions};
\draw[track] (\barleft,-1.86-\barh/2) rectangle (\barw,-1.86+\barh/2);
\draw[pressure] (\barleft,-1.86-\barh/2) rectangle (.72,-1.86+\barh/2);
\node[valuelabel] at (3.05,-1.86) {45+16 / 240};
\node[zerobox] at (\zerox,-1.86) {0};
\end{tikzpicture}
\caption{Proposal pressure before admission. Bars show unauthorized proposals
before the CXI gate; right-hand boxes show executed escapes at the mediated
sink. The figure measures model proposal pressure; backend evidence is
evaluated separately. Rows use different denominators and are not averaged.}
\label{fig:proposal-pressure}
\end{figure}
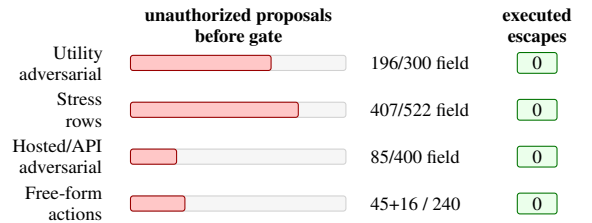

Figure~\ref{fig:proposal-pressure} summarizes the pressure rows as a
pre-gate measurement. Each row keeps its own denominator; the point is not an
average, but the contrast between unsafe proposal behavior before admission and
zero observed executed escapes at the mediated sink. The rows are 196/300
unauthorized field proposals for the utility adversarial setting, 407/522
unauthorized field proposals for stress rows, 85/400 unauthorized field
proposals for hosted/API adversarial rows, and 45 unauthorized field plus 16
unauthorized invocation proposals over 240 free-form actions.

\noindent\textbf{Choice pressure.}
Models still choose unauthorized fields when attack cases offer them, which is
why CXI enforces authority at the sink. In the proposal-before-gate experiments,
the model chooses between an authorized typed or opaque action and an
unauthorized raw-W action; candidate identifiers are opaque, and the gate runs
only after the choice. These rows measure proposal pressure; backend evidence
is evaluated separately. The field-policy prompt reduces the open-weight utility
adversarial subset to 67/300, while the hosted field-policy prompt produces
0/400 in the archived run. Call-level proposal pressure is measured separately.

\noindent\textbf{Free-form construction.}
Two free-form construction assays build complete action objects from task text
rather than choosing between opaque candidate IDs. The 60-case deterministic
free-form construction assay records 40 unauthorized field proposals, no
unauthorized invocation proposals, no executed field or invocation escapes, and
20 benign executions. The open-weight free-form run emits 240 free-form action objects
across the four open-weight models. All
240 are parseable; the runs record 45 unauthorized field proposals, 16
unauthorized invocation proposals, no field or invocation escapes reaching
execution, and 60 benign executions. GPT-OSS emits \texttt{call=false} for all
20 benign cases in this policy-aware free-form setting. Benign-call outcomes
are reported as availability metrics.

The takeaway is narrow: unauthorized proposal pressure appears before
enforcement and varies by model and prompting condition; decoder dependencies,
backend evidence, and task quality are evaluated at other boundaries.

\subsection{Schemas and Summaries Do Not Carry Authority}

\noindent\textbf{Schema baseline.}
Schema validation checks form, not authority. Across all 192 CXI-Core cases, a
raw co-resident context can read W and choose side effects directly: W-derived X
is 1.000, malicious-sink use is 1.000, benign admission is 0.000, and
unauthorized execution is 1.000. Schema-only structured output constrains JSON
shape but still allows W-derived values to fill X fields: W-derived X remains
1.000, malicious-sink use falls to 0.000, benign admission rises to 1.000, and
unauthorized execution remains 1.000. The \wxr boundary checks source authority
instead: the field must carry trusted authority or a destination-scoped typed release. Its
192-case row reports W-derived X=0.000, malicious-sink use=0.000, benign
admission=1.000, and unauthorized execution/gate escape=0.000.

The same evidence can receive different decisions in different fields. A
W-derived \texttt{approved} string is rejected for
\texttt{approval\_state}; a W-derived path validated as \texttt{FilePath} is
admitted only for \texttt{arguments.file\_path}; and a PR body may carry opaque
evidence but re-enters as W if a later bot or agent uses it to create side
effects. Free-form summaries fail for the same reason: ``approved'' remains
W-derived unless an authorized approval declassifier validates a trusted
approval reference.

\noindent\textbf{Mechanism baselines.}
We also include a mechanism-level baseline suite beyond schema checks. These
rows isolate which authority checks are lost without CXI's sink-and-field
specialization; they are not implementation-level comparisons with PACT,
CaMeL, FIDES/PFI, or ARGUS.
``Argument-role provenance approximation'' rejects 12/20 attack cases and
preserves 6/6 utility cases but leaves call-event and backend-binding failures;
``field-scoped capabilities, no call ledger'' rejects 10/20 attacks but leaves
invocation-authority failures; and ``typed reader/writer plus call ledger, no
field gate'' rejects 8/20 attacks but leaves cross-field typed-release reuse,
evidence record substitution, opaque reentry, and effect-bearing payloads. The
CXI row rejects 20/20 attack cases while preserving 6/6 utility cases.

\subsection{Why the CXI Mechanisms Are Non-Interchangeable}

The mechanisms in \wxr are not interchangeable. Transitive provenance blocks
summary and memory laundering; destination scope blocks one-field releases from
authorizing another field; semantic validators constrain typed outputs; opaque
reentry keeps copied evidence from becoming safe forever. Backend evidence
binding blocks replay, mutation, and schema-repair substitution. The distilled
map is: shape-valid laundering still needs a field-local gate; summary
laundering needs transitive provenance; wrong-field D reuse needs destination
scope; W-controlled retry or batching needs capability consume; opaque reentry
remains W; effect-bearing payloads need exact-effect authorization; and stale or
substituted backend records need manifest-bound evidence binding.

\noindent\textbf{Focused witnesses.}
Focused witness suites cover reviewer-raised corner cases without model
sampling. Field-authority witnesses exercise wrong trusted authority,
cross-field D reuse, value-digest mismatch, action-context mismatch, closed
optional-schema branches, and unmediated alternate sinks. Invocation-authority
witnesses exercise W-controlled retries, batch cardinality, sequence order,
idempotency keys, schedule time, missing invocation capabilities, ambiguous
capability identifiers, and the requirement that an accepting call consumes the
same selected capability identifier and receives an execution lease. Semantic
sink-binding witnesses check that the authorized object is the object consumed
after parser, canonicalizer, trusted-snapshot, renderer, and resolver effects.
These tests check distinct failure modes; aggregate ablations are reported
separately. Field-authority
binding covers 8 cases with 7 field proposals, no executed field or call
escapes, and 1 accept; invocation-authority binding covers 34 cases with 25
call proposals, no executed escapes, and 9 accepts; semantic sink binding
covers 8 cases with 6 field proposals, no executed escapes, and 2 accepts.

\noindent\textbf{Ledger witnesses.}
Ledger witnesses exercise admission separately from external-effect
multiplicity. Table~\ref{tab:main-results} reports the manifest-bound
rerun: 10 cases, 29 consume attempts, 19 rejections, 2 deduplicated retries, no
duplicate admissions, and no duplicate external applications. Additional
concurrent stress and crash/recovery campaigns also record no double-spends
under the declared idempotent-adapter contract. This is duplicate
suppression under the adapter contract, not a general exactly-once
external-effect claim.

Typed releases are the utility mechanism: removing them leaves the gate
conservative, but useful W-derived facts can no longer cross into protected
fields. These tests show distinct failure paths; they are not prevalence
estimates for arbitrary deployments.

\subsection{Boundary Utility and Task Quality}

\noindent\textbf{Boundary utility.}
\wxr keeps two utility channels open: typed releases carry narrow
facts for declared fields, and opaque data slots preserve evidence in comments,
tickets, PR bodies, or reports without granting authority to that text. The
same W-derived values are rejected when reused as \texttt{tool},
\texttt{operation}, or \texttt{approval\_state}.

We run 100 workflows over 270 boundary steps so utility is measured across
complete boundary steps, not isolated fields. The suite covers 50 code-agent
workflows over 150 steps, 20 Web/RAG workflows over 60 steps, 12 productivity
workflows over 24 steps, 10 ops workflows over 20 steps, and 8 memory workflows
over 16 steps. All declared workflows succeed at the boundary with no observed
false blocks: typed values are admitted only at destination scope, opaque
payloads are preserved, and matching raw-W probes are rejected for the same
field. This measures boundary utility, not end-to-end task success. The
remaining utility risks are practical policy risks: underspecified scopes,
over-narrow validators, missing opaque slots, and effect-bearing payloads
classified as ordinary data.

\noindent\textbf{Validator assurance.}
Validator assurance is field-dependent. The policy-surface audit
separates syntactic validators, trusted-lookup validators, and
semantic/effect-bounding validators; it records 20 D types over 21 audited
occurrences, 20 sinks, and 134 tests. A deterministic sandboxed code-agent
check over issue, README, CI-log, package-metadata, tool-output, and retry
risks shows why this matters: across 48 episodes, no-gate and schema-only
conditions admit all six unsafe actions; the argument-role provenance
approximation leaves one patch-payload gap and one retry gap; CXI blocks the
unsafe actions and preserves the six clean actions. The same harness checks
28/28 exact-effect authorization cases over repository state-delta effects.
This sandbox/oracle check is reported separately from live LLM benchmarks.

\noindent\textbf{Task-quality audit.}
We also run a 40-task final-state audit to keep task utility separate
from boundary utility. All 40 representative tasks complete the boundary, 29
satisfy task-level assertions without follow-up review, and the audit records
184/195 assertion passes. The 11 review-required cases are ordinary
task-quality issues such as missing regression tests, install-hook review,
ambiguous current-source evidence, approval/delegation scope review, namespace
review, reentry scope review, and stale status review. Task success still
depends on tests, review, and application-specific quality checks.

\subsection{Cost and Policy Effort}

The cost rows report boundary overhead rather than production throughput. The open-weight
CXI-Core evidence record path reports wall seconds per case plus peak CUDA
reserved memory; mask construction is about 0.16 ms per case at the median
(0.04 ms per protected field), so model execution and loading dominate.
Hosted/API rows are slower because each case uses structured reader and writer
round trips; full-request API latency ranges are 9.328--15.275 seconds for
AgentDojo hosted rows and 10.672--14.017 seconds for ToolEmu hosted rows.

The backend evidence footprint is about 677 bytes per case for four protected
fields. This cost measurement excludes vLLM worker peak memory because the
process-local CUDA snapshot does not capture worker VRAM. Canonical JSON
parseability is reported separately from backend evidence validity.

\noindent\textbf{Policy effort.}
CXI also exposes the policy work required to mediate representative sinks. This
policy-surface audit covers 24 evaluated
sinks: 149 fields, 85 X fields, 20 D fields, 29 O fields, 19 effect-bearing
payloads, and 141 validator tests. The initial 18 findings include missing
effect-bearing payloads, over-broad opaque slots, missing destination scopes,
natural-language approval, unmarked reentry, and validator scope gaps. The
regenerated audit records 9 high-severity and 9 medium-severity findings, no
unclassified findings, and no final blockers for the evaluated policy set.
This supports the evaluated policies; independently written policies still need
review.

%% file: sections/06_discussion_limitations.tex
\section{Discussion}

\paragraph*{Execution is the boundary.}
CXI evaluates security at the moment a candidate action becomes an external
effect. Unsafe proposals are attack pressure; the relevant measure is whether
an unauthorized field, effect, or invocation reaches a mediated sink. The
AgentDojo and code-agent rows support this admission claim under fixed
policies, snapshots, adapters, and task banks. Validator completeness, task
quality, and unreviewed sinks remain separate concerns.

\paragraph*{Authority is local.}
A log line may be quoted as opaque evidence, a path extracted from it may fill
one \texttt{file\_path} field after a typed release, and the same text still
has no authority for \texttt{operation}, \texttt{approval\_state},
\texttt{ci\_target}, or retry state. Effect-bearing payloads need
adapter-specific authorization because the authorized object is the
sink-interpreted effect, not the payload spelling. CXI does not ask whether
text is generally trustworthy, but whether a specific value, effect, or
invocation event carries authority for this sink manifest.

\paragraph*{Policy remains trusted work.}
Every mediated sink needs field classification, field-scoped declassifiers,
exact-effect validators, opaque slots, bypass tests, and complete mediation.
Broad validators, wrong field classes, unmediated sinks, or compromised
runtimes still break the guarantee; CXI makes that work explicit and does not
infer missing policy. The policy author must decide which fields select
authority, which payloads are interpreted by the sink, and which alternate
entry points could bypass the gate. The appendix summarizes the audited sink
families, bypass checks, adapter checks, and residual-risk examples.

\paragraph*{Evidence differs by regime.}
Open-weight deployments can verify protected-field decoding with backend
evidence records that bind each field to a single manifest before admission.
Hosted/API deployments cannot see provider internals, so they use the same
admission rule with host-observed compatibility evidence instead. That means
hosted/API checks can confirm the action fits the gate, but they do not prove
provider-internal enforcement for masks, attention, or KV-cache lineage.
Remote attestation against a compromised host and coverage for each transformer
implementation still need separate mechanisms; open-weight serving checks can
exercise the actual serving path.

\paragraph*{Takeaway.}
The model proposes actions, but the host admits only manifest-bound actions
that tie protected fields, interpreted effects, and the invocation event to
the same sink authority. A field, payload effect, or retry event is admitted
only when the corresponding authority object is present and valid. Unsafe text
may shape a proposal, but it only becomes authority through a typed release,
exact-effect authorization, or invocation capability.

%% file: sections/07_related_work.tex
\section{Related Work}

\paragraph*{Argument provenance and agent IFC.}
CXI's novel contribution is not provenance, IFC, nor capabilities; it
specializes them to manifest-bound admission of mixed-authority tool actions.
PACT identifies the granularity mismatch in agent security: whole-call trust is
too coarse when only some arguments carry authority~\cite{pact2026}. CXI targets
admission: plausible arguments may still lack authority to occur, recur, batch,
schedule, consume budget, or spend an idempotency token. Classical IFC supplies
labels, integrity, noninterference, and declassification~\cite{denning1976lattice,goguen1982security,
volpano1996sound,sabelfeld2003language,myers1999jflow,myers2000dlm,
efstathopoulos2005asbestos,zeldovich2006histar,krohn2007flume,
sabelfeld2009declassification}; FIDES, Prompt Flow Integrity, and FORGE bring
related ideas to LLM agents and policy enforcement~\cite{fides,pfi,forge2026}.
CXI uses this lineage narrowly: it records enough field, effect, and invocation
evidence for the host to reject one concrete sink call when the authority chain
does not close.

\paragraph*{Action authority and capabilities.}
ARGUS and AIRGuard also check action authority near execution~\cite{
argus2026,airguard2026}. CXI's contribution is a deterministic predicate over
a canonical sink object: verify protected fields, consume the matching
invocation capability, and execute only with the resulting lease. This follows
the confused-deputy and capability tradition~\cite{hardy1988confused,
levy1984capability,shapiro1999eros,watson2010capsicum,watson2015cheri,
wahbe1993sfi,yee2009nativeclient,abadi2005cfi}. CaMeL extracts control and data
flows from a trusted query and executes the resulting program with
capabilities~\cite{camel}; CXI operates later, checking candidate actions at
admission time. A candidate action can be syntactically valid, policy-shaped,
and useful while still lacking authority for the field, interpreted effect, or
invocation event that would make it execute.

\paragraph*{Prompt defenses, serving stacks, and benchmarks.}
Prompt-injection defenses, instruction hierarchies, StruQ, Prompt Control-Flow
Integrity, guards, judges, sanitizers, and priority-aware middleware reduce
unsafe proposals or prompt-flow confusion~\cite{greshake2023indirect,
perez2022ignore,struq,wallace2024instruction,pcfi,zou2023universal,
wei2023jailbroken,chao2023pair,mehrotra2023tree}. CXI assumes unsafe proposals
may still occur and checks authority before execution.
High-throughput serving stacks introduce KV-cache reuse, batching, custom
attention kernels, guided decoding, and fast paths~\cite{vllm,
dao2022flashattention,dao2023flashattention2,orca,flexgen,sglang,
speculative-decoding}; CXI requires backend evidence for supported paths and
fail-closed behavior for unsupported ones. Tool-use benchmarks provide task
environments~\cite{agentdojo,toolemu,swebench,webarena,osworld,agentbench,
appworld,taubench,browsergym}; CXI keeps benchmark utility separate from
admission integrity.

%% file: sections/08_conclusion.tex
\section{Conclusion}

LLM agents read untrusted text and act through real tools with external
effects. CXI makes that transition explicit by admitting only actions whose
field evidence, exact-effect authorization, and invocation capability all bind
to the same canonical manifest before the mediated sink executes. It makes the
action admissible only when the gated authority chain is complete.

Untrusted evidence may still be quoted, summarized, validated into a typed
release, or bound to an authorized effect. Those uses preserve provenance, but
they do not silently grant authority for a different field, effect, or
invocation event.

This is a concrete systems property for mediated sinks. It does not make
validators complete, tasks correct, provider internals observable, or external
effects exactly-once on its own. What it does do is make the final step from
context to execution explicit, checkable, and fail-closed: before the host
grants execution, the fields, effects, and invocation event must each point to
where their authority came from.

%% file: sections/08_ethics_open_science.tex
\section*{Ethics Considerations}

This work evaluates a boundary for side-effect authority in tool-using agents.
The experiments use public benchmarks, synthetic action-selection cases,
sandboxed repositories, and archived API traces; they do not access real user
accounts, secrets, payments, or production services. Attack cases are evaluated
as candidate actions, gate decisions, and sandboxed effects rather than
dangerous live side effects.

The main dual-use risk is that failure examples may help attackers reason about
prompt-injection and authority-laundering patterns. The release therefore
emphasizes policies, declassifier contracts, gate decisions, evidence records,
and aggregate results rather than operational playbooks against live systems.

\section*{Open Science}

An anonymous review artifact is available at
\url{https://anonymous.4open.science/r/cxi}. The primary entry point is
\texttt{submission\_artifact/}. The release contains the operational audit
material for the paper: policy and tool-schema manifests, field
classifications, declassifier specifications and negative tests, gate decisions,
backend evidence records, manifest-bound ledger traces, benchmark mappings,
frozen result records, generated tables, checksums, and aggregate-table
regeneration scripts. The default audit path regenerates the reported tables
and proposal-pressure figure from checked-in records and requires no provider
API keys, model weights, GPU access, or production service access.

Run manifests tie sampled rows to model settings, policy hashes, source hashes,
and result hashes so compatible serving stacks can replay the backend-evidence
checks. Live open-weight and hosted/API reruns are documented as optional reruns
rather than required review steps.

Hosted/API results are archived compatibility traces rather than
provider-internal evidence. They preserve requested model identifiers, returned
identifiers where available, timestamps, provider settings, request/response
traces, selections, gate decisions, and boundary evidence, but cannot expose
provider-internal attention, masks, or KV-cache lineage. Released records exclude
credentials, local machine addresses, local paths, and author-identifying
metadata.

%% file: sections/09_appendix.tex
\section*{Appendices}
\renewcommand{\arraystretch}{0.94}

The appendices provide technical details for readers who want to inspect the
admission boundary behind the main text. Appendix~\ref{app:boundary-obligations}
states the boundary obligations; Appendix~\ref{app:security-contract-details}
gives the operational contract and gate checks;
Appendices~\ref{app:deployment-discussion-checklists}--\ref{app:mechanism-coverage}
cover complete mediation, typed releases, exact effects, and mechanism
composition; Appendix~\ref{app:generated-analysis-traceability} records the
evaluation setup and model settings;
Appendices~\ref{app:evidence-regimes}--\ref{app:result-row-backing} describe
evidence boundaries and the records behind the main result rows; and
Appendix~\ref{app:residual-risks} summarizes residual deployment risks.

\section{Boundary Obligations at a Glance}
\label{app:evaluation-matrix}
\label{app:boundary-obligations}

Table~\ref{tab:boundary-obligations} summarizes the obligations checked before
an action becomes executable.

\begin{table}[!bp]
\caption{Boundary obligations. Each obligation binds one part of the execution
boundary to the same canonical action manifest.}
\label{tab:boundary-obligations}
\scriptsize
\begin{tabularx}{\columnwidth}{p{0.24\columnwidth}>{\raggedright\arraybackslash}X>{\raggedright\arraybackslash}X}
\toprule
Obligation & What must be bound & Typical failure blocked \\
\midrule
Field authority & Protected sink field value, destination scope, policy epoch,
trusted snapshot, and action context. & Copied approval text, wrong-field typed
release, stale trusted state, or raw W selecting a protected field. \\
Effect authority & Sink-parsed payload effect under the trusted snapshot,
adapter revision, principal, and manifest digest. & Same patch on another base,
path rewrite, SQL role mismatch, or raw command text from W. \\
Invocation authority & Call, retry, batch, schedule, or delegation event plus a
consumable invocation capability. & Stale capability, duplicate consume,
wrong-manifest replay, or W-controlled retry. \\
Complete mediation & Every side-effecting route reaches the shared gate before
execution. & Helper API, alternate client, generated file, bot, or install hook
bypassing admission. \\
Evidence binding & Field evidence, exact-effect record, ledger consume, policy
epoch, snapshot, backend/API profile, and parent records. & Evidence
substitution, lower evidence mode, repair after evidence, or stale parent
record. \\
\bottomrule
\end{tabularx}
\end{table}

\section{Operational Contract and Gate Checks}
\label{app:security-contract-details}
\label{app:field-authorization-predicate}
\label{app:sink-boundary-verifier}

Let an action manifest be
$m=(sink,op,\rho,\nu,\sigma,\eta,\vec f,\vec p)$, where $\rho$ is the task/run
context, $\nu$ the policy epoch, $\sigma$ the trusted sink snapshot, $\eta$ the
action nonce, $\vec f$ the protected field assignments, and $\vec p$ the
sink-interpreted payloads. The gate receives policy $P$, field provenance
$\Pi[f]$, call provenance $\Pi_{\mathrm{call}}$, effect commitments $E$, and
required backend or field-local evidence records $B$.

Field atoms have four forms. $T(p,\alpha,\sigma,b,\rho,\nu)$ is trusted
authority in domain $p$. $W(o)$ is writable influence. $D(\tau,\sigma_W,
\sigma_X,d,v,b,\kappa,\rho,\nu,I)$ is a typed release from W-scope $\sigma_W$
to destination $\sigma_X$, minted by declassifier $d$ after validator $v$
accepts binding $b$ under action constraint $\kappa$. $O(h,I)$ is an opaque data
handle with parent influence $I$. Call atoms use the same shape but carry a
capability identifier. Unknown, stale, or malformed atoms reject.

\begin{table}[!htbp]
\caption{Operational influence events. CXI tracks what may have selected a
value, not only where its bytes came from. Derived values keep W influence until
trusted code mints a destination-scoped typed release.}
\label{tab:provenance-events}
\scriptsize
\begin{tabularx}{\columnwidth}{p{0.24\columnwidth}>{\raggedright\arraybackslash}X>{\raggedright\arraybackslash}X}
\toprule
Event & Influence rule & Fail-closed condition \\
\midrule
Read trusted state & Adds field- or call-scoped $T$. & Missing policy, wrong
domain, stale snapshot, or untrusted origin metadata. \\
Read writable context & Adds $W(o)$ for attacker-writable or attacker-influenced
origin $o$. & Unknown origin is treated as W. \\
Summarize/copy text & Union of all visible inputs. & Summary later used as
authority without D. \\
Extract typed fact & Remains W until validator emits D. & Type, scope,
destination, validator, or policy mismatch. \\
Memory or tool output & Preserves prior influence; tool output is W unless
policy marks it trusted. & Missing provenance or tool output selects a protected
field without D. \\
Repair/retry/batch & Union of candidate, repair context, previous action, and
control inputs. & Protected field or event changes without fresh authority. \\
Opaque data / reentry & Opaque data carries parent influence but no current
authority. & Later side-effect boundary treats reentry as W. \\
\bottomrule
\end{tabularx}
\end{table}

\begin{table}[!htbp]
\caption{Gate checks. Each check is evaluated against one canonical manifest,
policy epoch, trusted snapshot, and action context.}
\label{tab:admission-predicates}
\scriptsize
\begin{tabularx}{\columnwidth}{p{0.25\columnwidth}>{\raggedright\arraybackslash}X>{\raggedright\arraybackslash}X}
\toprule
Predicate & Question & Rejects \\
\midrule
\textsc{MayInfluence} & May this atom participate in choosing this field or
event? & Wrong sink, field, operation, authority domain, destination scope,
policy epoch, or action context. \\
\textsc{AuthorizesValue} & Does this atom authorize the executed semantic value
or effect? & Digest mismatch, stale snapshot, wrong adapter, wrong validator,
wrong principal, or cross-field D reuse. \\
\textsc{EffectOK} & Does an adapter-specific commitment authorize the
sink-interpreted payload? & Payload replay, post-validation mutation, wrong base
state, stale policy, or untrusted renderer. \\
\textsc{AuthorizeInvoke} & May this call, retry, batch, schedule, or delegation
event occur? & No capability, wrong manifest, stale sequence, budget mismatch,
or duplicate spend. \\
\bottomrule
\end{tabularx}
\end{table}

For a protected field $f$ with value $v_f$, $\mathrm{FieldOK}(f)$ holds when
$f$ appears in the active sink schema, $P$ classifies $f$, $\Pi[f]$ is known and
fresh, every atom in $\Pi[f]$ may influence $f$, and at least one trusted or D
atom authorizes the canonical value. Raw W, opaque-only O, cross-field D,
wrong-domain T, stale policy versions, repaired protected fields, and missing
referenced fields reject. For each sink-interpreted payload,
$\mathrm{EffectOK}$ requires an exact-effect authorization over the effect that
the sink will apply under the trusted snapshot, policy epoch, principal, adapter
revision, and action nonce. For the invocation event, every atom in
$\Pi_{\mathrm{call}}$ must be allowed by invocation policy. The resolver must
identify exactly one capability to spend, unless policy declares a deterministic
equivalence or priority rule. The selected capability is consumed by a
linearizable ledger against $digest(m)$; only the returned lease may reach the
sink.

\noindent\textbf{Verifier skeleton.}
The gate first canonicalizes the candidate under the trusted snapshot and
rejects malformed, policy-incompatible, or post-validation-mutated manifests.
It then checks each protected field with \textsc{MayInfluence} and
\textsc{AuthorizesValue}, verifies required backend or field-local evidence for
that field, checks exact-effect commitments for each sink-interpreted payload,
resolves one invocation capability, consumes that capability against the digest
of the same manifest, and sends only the resulting lease to the mediated sink.
The admitted-action contract is therefore
\[
  \mathrm{Admit}(m)\Rightarrow
  \mathrm{Fields}(m)\land\mathrm{Effects}(m)\land\mathrm{Invoke}(m).
\]
\noindent Here $\mathrm{Fields}(m)$ means all protected fields satisfy the field
predicate; $\mathrm{Effects}(m)$ means all sink-interpreted payloads match
exact-effect commitments for the same manifest; and $\mathrm{Invoke}(m)$ means a
manifest-bound capability is consumed into an execution lease. The trusted
assumptions are complete mediation, conservative influence tracking, correct
policy classification, fail-closed validators, and evidence records that
over-approximate influence.

\section{Complete Mediation Checks}
\label{app:deployment-discussion-checklists}

These entries are part of the claim, not optional hardening. If a deployment
leaves a side-effecting route outside the mediated path, CXI no longer claims
admission integrity for that route. The gate can reject an unauthorized
manifest; it cannot protect a sink call that bypasses it.

\begin{table}[!htbp]
\caption{Side-effect families and bypass checks. The gate protects a deployment
only when every side-effect path reaches it.}
\label{tab:main-mediation-app}
\scriptsize
\begin{tabularx}{\columnwidth}{p{0.20\columnwidth}>{\raggedright\arraybackslash}X>{\raggedright\arraybackslash}X}
\toprule
Sink family & Mediated entry points and authority & Negative bypass checks \\
\midrule
Filesystem & Write, delete, move, generated files; path, operation, and payload
effect require authority. & Direct file APIs and helper paths are denied or
wrapped. \\
Shell/process & Trusted-rendered argv/env/cwd plus sandbox, network, and
filesystem policy. & Raw shell strings from W do not execute. \\
Network/API & Endpoint, recipient, namespace, method, and idempotency bind to
the manifest. & Alternate clients require the same policy and lease. \\
Repository & Patch, branch, PR, workflow, and CI changes bind path and state
delta. & Patch bytes cannot execute raw or replay on another base. \\
Memory/state & Store, retrieve, summarize, and repair preserve provenance. &
Unlabeled or stale memory fails closed. \\
Delegation/bots & Agent target, task, permissions, and event capability are
protected. & Opaque data reentry becomes W. \\
Package/CI & Dependency, registry, workflow, and install hooks use exact-effect
validators. & Install-hook and workflow side effects require adapter authority. \\
\bottomrule
\end{tabularx}
\end{table}

\section{Typed Release and Exact-Effect Contracts}
\label{app:implementation-tables}

\begin{table}[!htbp]
\caption{Release and effect contracts. A typed release grants one
destination-scoped value; exact-effect authorization grants the sink-interpreted
effect under the trusted snapshot.}
\label{tab:validator-examples}
\scriptsize
\begin{tabularx}{\columnwidth}{p{0.25\columnwidth}>{\raggedright\arraybackslash}X>{\raggedright\arraybackslash}X}
\toprule
Mechanism & Binding checked & Failure blocked \\
\midrule
File path release & Repository-relative path for one destination field after
canonicalization and scope check. & Tool choice, CI target, shell command, or
package target. \\
Approval release & Trusted approval reference bound to principal, epoch, and
action context. & Copied ``approved'' text. \\
Dependency release & Registry-scoped dependency name for one dependency field.
& Registry switch or install hook. \\
Patch intent release & State-delta authorization under the trusted repository
snapshot. & Arbitrary patch bytes or replay on another base. \\
Repository patch adapter & Base commit, canonical paths, old/new blobs, modes,
renames, and allowed path set. & Patch replay, path switch, stale base, or
post-validation mutation. \\
File body / SQL / shell adapters & Sink-parsed payload, role/snapshot, trusted
renderer, and sandbox or endpoint policy. & Raw string authority, role mismatch,
wrong path, command-text laundering, or sandbox escape. \\
\bottomrule
\end{tabularx}
\end{table}

These contracts separate value authority from effect authority. Before the gate
treats a release, effect commitment, or capability as authority, the validator
must bind destination scope, sink-parsed value, trusted snapshot,
post-validation immutability, adapter coverage for the sink class, and the
mediated entry point. A typed release can justify one field value, but it does
not authorize the side effect of interpreting a patch, SQL statement, workflow,
or command. The adapter checks close that second step under the trusted
snapshot.

\section{Mechanism Coverage and Composition}
\label{app:mechanism-coverage}

\begin{table}[!htbp]
\caption{Mechanism coverage. Different failures need different boundary checks;
no single partial mechanism covers all rows.}
\label{tab:mechanism-coverage-app}
\scriptsize
\begin{tabularx}{\columnwidth}{p{0.36\columnwidth}>{\raggedright\arraybackslash}X}
\toprule
Failure class & Boundary component needed \\
\midrule
Shape-valid authority laundering & Field-local gate over protected sink fields. \\
Summary or memory laundering & Transitive influence provenance. \\
Wrong-field D reuse & Destination-scoped typed release. \\
W-controlled retry/batch/schedule & Invocation capability plus consume. \\
Opaque reentry & Opaque data reenters later boundaries as W. \\
Effect-bearing payload & Exact-effect authorization. \\
Evidence substitution or repair & Manifest-bound evidence record. \\
Stale policy or snapshot & Policy epoch and trusted snapshot digest. \\
Unmediated alternate sink & Complete mediation of side-effect paths. \\
\bottomrule
\end{tabularx}
\end{table}

The rows are independent failure paths. A schema can rule out malformed
objects, but it does not say who chose a protected field. A call ledger can
prevent duplicate invocation, but it does not authorize patch content or
approval state. A typed release can carry one validated value, but it cannot be
reused for another field. CXI closes these cases only when the field, effect,
and invocation checks meet at the same manifest.

\section{Evaluation Setup and Model Settings}
\label{app:generated-analysis-traceability}

The setup records identify where sampled evidence was collected. They are not
result categories. Deterministic rows use fixed checks and fault-injection
cases; sampled rows use the model and serving configurations below. Hosted/API
aliases are archived record identifiers and may not denote provider-internal
model revisions.

\begin{table}[!htbp]
\caption{Model/runtime/settings matrix. Identifiers and settings are copied
from recorded run configurations; ``not sent'' means deliberately omitted.}
\label{tab:model-runtime-matrix}
\scriptsize
\begin{tabularx}{\columnwidth}{p{0.42\columnwidth}>{\raggedright\arraybackslash}X}
\toprule
Path and model identifiers & Recorded setup \\
\midrule
Open-weight vLLM BF16: \path{Qwen/Qwen3.6-27B};
\path{google/gemma-3-27b-it};
\path{Qwen/Qwen3-Coder-30B-}\newline\path{A3B-Instruct}. &
\texttt{dtype=bfloat16}; \texttt{max\_model\_len=2048};
\texttt{max\_new\_tokens=64}; \texttt{temperature=0.0};
\texttt{gpu\_memory\_utilization=0.85}; \texttt{max\_num\_seqs=1}. \\
GPT-OSS vLLM MXFP4: \path{openai/gpt-oss-120b}. &
\texttt{dtype=auto}; \texttt{quantization=gpt\_oss\_mxfp4};
\texttt{max\_model\_len=1024}; \texttt{max\_new\_tokens=64};
\texttt{gpu\_memory\_utilization=0.96}. \\
OpenAI hosted: \path{gpt-5.5}. &
\texttt{chat.completions}; strict JSON schema response format;
\texttt{max\_completion\_tokens=2048}; \texttt{seed=0};
\texttt{temperature=1.0}; \texttt{top\_p=1.0};
\texttt{reasoning\_effort=medium}; four retries; 120s timeout. \\
Anthropic hosted: \path{claude-opus-4-7}; \path{claude-sonnet-4-6}. &
\texttt{messages}; \texttt{anthropic\_version=2023-06-01};
\texttt{max\_tokens=2048}; forced tool choice; seed unsupported;
temperature and top-p not sent; four retries; 120s timeout. \\
Gemini hosted: \path{gemini-3.5-flash}. &
\texttt{chat.completions}; strict JSON schema response format;
\texttt{max\_completion\_tokens=2048}; seed not sent;
\texttt{temperature=0.0}; \texttt{top\_p=1.0}; four retries; 120s timeout. \\
\bottomrule
\end{tabularx}
\end{table}

Open-weight sampled rows ran on NVIDIA H100 80GB HBM3 GPU hosts with driver
570.172.08, CUDA 12.8, PyTorch 2.11.0+cu128, and vLLM 0.21.1rc1.dev268
precompiled builds. GPT-OSS used native MXFP4. This specifies the machine and
serving environment; it does not define a separate evidence claim.

\section{Evidence Boundaries}
\label{app:evidence-regimes}

Different evidence paths support different parts of the same admission
predicate. The boundaries below say what each path contributes before the gate
and what remains outside the claim.

\begin{table}[!htbp]
\caption{Evidence boundary summary. Each path supports a specific part of the
admission claim and leaves different non-claims outside the boundary.}
\label{tab:run-lineage-summary}
\scriptsize
\begin{tabularx}{\columnwidth}{p{0.22\columnwidth}>{\raggedright\arraybackslash}X>{\raggedright\arraybackslash}X}
\toprule
Path & Supports & Does not support \\
\midrule
Field-closed open-weight & Reference target for protected-field decoding:
fresh projection, field grammar, parent records, nonce, backend profile. &
All serving stacks or compromised-host attestation. \\
Direct-span / suffix-taint / runtime-map & Declared raw-W source,
generated-prefix, or declared-atom exclusion before field admission. &
Full hidden-state, KV-lineage, or model-level noninterference. \\
Open-weight serving check & Field-local construction and evidence binding on
the evaluated open-weight profile. & Other optimized backend variants unless
they satisfy the same gate check. \\
Hosted/API compatibility & Reader/extractor and field-local writer visibility
when provider internals are unavailable. & Provider-internal masks, attention,
KV-cache lineage, or enforcement. \\
Exact-effect adapters & Sink-interpreted state deltas or commitments for
effect-bearing payloads. & Validator completeness beyond the adapter contract. \\
Ledger & One manifest-bound capability consume and execution lease. & Automatic
exactly-once external delivery without adapter idempotency. \\
\bottomrule
\end{tabularx}
\end{table}

\noindent\textbf{Common record invariants.}
All evidence verifiers bind the action digest, protected field path, policy
epoch, trusted snapshot, backend or hosted/API profile, action nonce, adapter
revision, and parent records. They reject field changes after evidence
recording, evidence for a different field, stale policy or snapshot, lower
evidence mode than policy requires, unsupported backend paths, hosted/API
writers that see raw W for protected fields, stale or wrong-manifest capability
consume, exact-effect replay, stale base states, and path rewrites. Missing
policy fields, broad validators, and unmediated sinks remain deployment risks
outside CXI admission.

\section{Result-Row Backing}
\label{app:result-row-backing}

The result records keep denominators separate. Episodes, field records, leases,
consume attempts, archived hosted/API rows, and proposal rows are different
units, so the paper never averages them into one success rate. A row supports
the admission claim only when the recorded action reaches the mediated sink with
field, effect, and invocation authority checked against the same manifest.

\begin{table}[!htbp]
\caption{Evidence behind the main result rows. The table names the checked
records behind each denominator without replaying individual traces.}
\label{tab:main-result-record-map}
\scriptsize
\begin{tabularx}{\columnwidth}{p{0.25\columnwidth}>{\raggedright\arraybackslash}X>{\raggedright\arraybackslash}X}
\toprule
Main item & Records checked & What the row establishes \\
\midrule
AgentDojo live task bank & Episodes, LLM calls, STC rows, protected-admission
outcomes. & 720 episodes, 1,739 calls, and 199 STC define the evaluated live
task surface. \\
AgentDojo exact-effect admissions & Field records, leases, effect
authorizations, gate decisions. & 306 field records and 204 leases/effects
support the exact-effect admission subset. \\
Code-agent exact-effect & Repository episodes, manifests, exact effects,
leases, task oracle outcomes. & 400 episodes, 231 STC, and the 73/61/35 non-STC
split separate parsing, authorization, and task quality. \\
Ledger faults & Consume attempts, rejects, dedup retries, admitted leases. &
10 cases, 29 consumes, 19 rejects, and 2 dedup retries support manifest-bound
consume/dedup. \\
Hosted/API compatibility & Archived action rows, field-local construction
records, and gate decisions. & 800 rows, 85 unsafe selections blocked, and 0
gate escapes support hosted/API compatibility. \\
Proposal pressure & Adversarial proposal rows and executed-escape counters. &
Figure 3 measures pre-gate unsafe proposal pressure separately from execution. \\
\bottomrule
\end{tabularx}
\end{table}

\section{Residual Risks and Deployment Checklist}
\label{app:residual-risks}

\noindent\textbf{Policy audit.}
Policy authoring is part of the trusted boundary, not an automatically inferred
property. The evaluated policy set records protected fields, typed-release
destinations, opaque data slots, effect-bearing payloads, validator tests, and
bypass checks for each mediated sink. The useful failure classes are practical:
a field can be left unprotected, a typed release can be scoped too broadly, an
opaque slot can reenter a downstream bot, an exact-effect adapter can miss a
sink-parsed effect, or a helper API can bypass the shared gate. CXI exposes
those cases as policy or mediation failures; it does not make the missing policy
true.

\noindent\textbf{Residual risks.}
The residual risks have the same shape as the contract assumptions. A
compromised host can lie about provenance or evidence records. A
provider-internal hosted decoder cannot expose masks, attention, or KV-cache
lineage to the reviewer. A validator can be too weak for a sink class, even
when the gate enforces it consistently. A task can fail for parser, benchmark,
or quality reasons after the admission decision is correct. These cases remain
outside mediated admission integrity; the claim stays narrow: field, effect,
and invocation authority must bind to the same canonical manifest before
execution.

\noindent\textbf{Deployment checklist.}
For a new mediated sink, the minimum checklist is small but strict. The policy
must classify every field that selects, authorizes, or parameterizes a side
effect. Each typed release must name its source scope, destination field,
validator, trusted snapshot, policy epoch, and action context. Each
effect-bearing payload needs an adapter that validates the sink-parsed effect,
not just the JSON shape or byte string. Each invocation capability must name the
same manifest digest that the field and effect evidence name, and the ledger
must consume it before execution. Alternate clients, helper APIs, retry paths,
batching, delegation, package hooks, and generated files must reach the same
gate or stay outside the claim. If a deployment cannot answer one of these
checks, the correct outcome is not a weaker claim about CXI; it is an unmediated
or under-specified sink.